\documentclass[12pt]{article}
\usepackage{docmute}

\usepackage{cite}
\usepackage{graphicx}
\usepackage{dcolumn}
\usepackage{bm}

\usepackage[hidelinks]{hyperref}
\makeatletter
\renewcommand{\theequation}{\thesection.\arabic{equation}}
\makeatother

\usepackage{epsf}
\usepackage{amsmath,amssymb}
\usepackage{slashed}
\usepackage{graphicx}
\usepackage{subfig}
\usepackage[usenames,dvipsnames]{xcolor}

\usepackage{tikz}
\usetikzlibrary{quantikz2}

\usepackage{physics}
\usepackage{dsfont}

\usepackage{comment}

\usepackage{tikz}
\usetikzlibrary{quantikz2}

\begin{document}

\begin{titlepage}

\def\thefootnote{\fnsymbol{footnote}}

\begin{center}

\hfill \today\\

\vskip .5in

{\Large \bf
  
  Dark Photon Dark Matter from Quantum Fluctuations during Starobinsky Inflation
  
}

\vskip .5in

{\large
  Taiyo Kasamaki, Takeo Moroi
}

\vskip .5in

{\em Department of Physics, The University of Tokyo, Tokyo 113-0033, Japan}

\end{center}
\vskip .5in

\begin{abstract}

We present a detailed investigation of scenarios in which dark-photon dark matter is produced from quantum fluctuations during inflation. In particular, we focus on inflationary models that necessarily involve a Weyl transformation, dependent on the inflaton amplitude, in order to move to the Einstein frame. In such models, the kinetic function of the longitudinal mode of the dark photon varies throughout, and even after, the inflationary period. We show that this variation of the kinetic function has a substantial impact on the resulting relic abundance of dark photons. As a representative and phenomenologically important example, we analyze the Starobinsky inflation model, for which we perform an accurate computation of the relic dark-photon abundance. By imposing the relevant observational constraints, we find that, in order to reproduce the observed dark-matter density in the present Universe, the dark-photon mass must lie in the range $5.6 < m < 7.4\,\mu\mathrm{eV}$ within the framework considered in this work.

\end{abstract}

\end{titlepage}

\renewcommand{\thepage}{\arabic{page}}
\setcounter{page}{1}
\renewcommand{\thefootnote}{\#\arabic{footnote}}
\setcounter{footnote}{0}
\renewcommand{\theequation}{\thesection.\arabic{equation}}

\section{Introduction}
\setcounter{equation}{0}

One of the central challenges in particle physics and cosmology is the identification of the dark matter (DM) component of the Universe. Astrophysical and cosmological observations have firmly established the existence of DM in our Universe. In particular, the present DM energy density has been determined with high precision; its density parameter is given by \cite{Planck:2018vyg}
\begin{align}
  \Omega_{\rm DM} h^2 = 0.120 \pm 0.001 \,,
\end{align}
where $h$ denotes the present-day dimensionless Hubble parameter. Despite this observational success, the particle-physics properties of DM remain largely unknown; in particular, its mass and the strength of its interactions with Standard Model (SM) particles are essentially unconstrained. Direct detection experiments are expected to provide crucial information on these properties, and substantial experimental efforts are currently under way.

A wide variety of DM candidates has been proposed and studied (for reviews, see, e.g., Refs.\ \cite{Feng:2010gw, Cirelli:2024ssz}). These candidates are often classified into two categories: particle-like DM and wave-like DM. Particle-like DM is characterized by a relatively large mass, which typically leads to a small occupation number within its de Broglie wavelength and, consequently, particle-like behavior. In contrast, wave-like DM has a sufficiently small mass that its occupation number becomes very large; as a result, such DM can often be described as a classical field obeying the classical equations of motion.

For wave-like DM, which is the main subject of this Letter, our ignorance of the DM mass presents an additional difficulty for many experimental searches, in particular haloscope-type experiments. Many haloscope experiments are designed to detect excitations in a quantum system that couples to the DM field; representative examples include electromagnetic modes (i.e., photons) in a resonant cavity and superconducting qubits. In order to enhance the excitation probability, the energy gap (or resonance frequency) of the quantum system must match the DM mass. To cover a wide range of possible DM masses, many experiments therefore scan the resonance frequency, which is typically technically demanding and time-consuming. Any theoretical or phenomenological information on the DM mass is thus valuable, as it can guide the choice of target frequency and inform an efficient scanning strategy.

In this work, we focus on one of the well-motivated candidates for wave-like DM, namely the dark photon. The dark photon is a massive vector field that kinetically mixes with the ordinary photon. As we describe below, an oscillating dark-photon field in the present Universe can constitute DM and is therefore an important target of haloscope searches. The dark-photon mass relevant for dark-photon DM, however, depends on the production mechanism and is consequently thought to be a free parameter. Among the various possibilities, an attractive production mechanism for dark-photon DM is based on quantum fluctuations of the dark-photon field generated during inflation \cite{Graham:2015rva}. (For more discussion about the production mechanisms of dark-photon DM, see Refs.\ \cite{Jaeckel:2008fi, Pospelov:2008jk, Redondo:2008ec, Nelson:2011sf, Arias:2012az, An:2013yfc, Redondo:2013lna, Tang:2017hvq, Garny:2017kha, Agrawal:2018vin, Dror:2018pdh, Co:2018lka, Bastero-Gil:2018uel, Long:2019lwl, Ema:2019yrd, Nakayama:2019rhg, Nakayama:2020rka, Nakai:2020cfw, Ahmed:2020fhc, Kolb:2020fwh, Salehian:2020asa, Firouzjahi:2020whk, Nakayama:2021avl, Wang:2022ojc, Sato:2022jya, Redi:2022zkt, Nakai:2022dni, Kitajima:2022lre, Gan:2023wnp, Kitajima:2023fun}.) In this scenario, the quantum fluctuations of the longitudinal mode during inflation can be sufficiently large to source the present DM abundance.

The main goal of this Letter is to provide a precise and detailed calculation of the DM abundance in this inflationary-fluctuation scenario and to determine the dark-photon mass relevant for dark-photon DM. As we will discuss, the subsequent evolution of the dark-photon amplitude generated during inflation depends on the cosmological history from the time the fluctuations are produced until the present epoch. For the case of simple single-field inflation, this evolution was studied in detail in Ref.\ \cite{Graham:2015rva}. However, in more general inflationary scenarios, properties of the inflaton may significantly modify the estimate presented in Ref.\ \cite{Graham:2015rva}. In particular, we focus on inflation models in which a Weyl transformation is inevitably required to reach the Einstein frame. In such cases, the kinetic function of the longitudinal mode of the dark photon depends on the inflaton field value, thereby affecting the amplitude of the dark-photon fluctuations. (For a study of a model with time-dependent kinetic function, see also Ref.\ \cite{Wang:2022ojc}.) This effect may substantially change the resulting DM abundance sourced by inflationary quantum fluctuations. Notably, Starobinsky inflation \cite{Starobinsky:1980te}, one of the leading inflationary models consistent with cosmological observations, provides an important and well-motivated example of this class. We will see that, in the case of Starobinsky inflation, the dark-photon mass should be within $5.6 < m < 7.4\,\mu\mathrm{eV}$ to realize the correct dark-matter abundance.

\section{Dark Photon Evolution}
\label{sec:evolution}
\setcounter{equation}{0}

We begin by outlining the basic framework for calculating the mass density of dark-photon DM produced in the class of inflationary models of interest. As mentioned above, we focus on inflationary models that require a Weyl transformation in order to move from the Jordan (non-Einstein) frame to the Einstein frame. We assume that, in the Jordan-frame Lagrangian, the mass term of the dark photon does not depend explicitly on the inflaton field (aside from its dependence through the metric); this is naturally realized in Starobinsky inflation.

Compared to the case of the original proposal~\cite{Graham:2015rva}, in our scenario, it is crucial to carefully consider the effect of the Weyl transformation from the Jordan frame to the Einstein frame. Hereafter, quantities defined in the Jordan frame are denoted with a ``bar.'' The metrics in the Jordan and Einstein frames, denoted as $\bar{g}_{\mu\nu}$ and $g_{\mu\nu}$, respectively, are related via
\begin{align}
  g_{\mu\nu} = \Omega^2 \bar{g}_{\mu\nu},
  \label{WeylTr}
\end{align}
where $\Omega$ is a function of the inflaton amplitude and is normalized such that
\begin{align}
  \Omega (t_{\rm now}) = 1,
\end{align}
with $t_{\rm now}$ being the present cosmic time.

Focusing on the dark-photon sector, the relevant part of the Lagrangian in the Jordan frame is
\begin{align}
  \mathcal{L}_{\rm DP} = - \frac{1}{4} \bar{g}^{\mu\nu} \bar{g}^{\rho\sigma}
  F_{\mu\rho} F_{\nu\sigma}
  + \frac{1}{2} m^2 \bar{g}^{\mu\nu} A_\mu A_\nu,
  \label{Lagrangian_J}
\end{align}
where $A_\mu\equiv (A_t, A_1, A_2, A_3)$ is the dark-photon field, $F_{\mu\nu}\equiv \partial_\mu A_\nu-\partial_\nu A_\mu$ is the field-strength tensor, and $m$ is a mass parameter. (With the normalization $\Omega (t_{\rm now})=1$, $m$ can be identified with the dark-photon mass in the present Universe.)

After performing the Weyl transformation, the action in the Einstein frame becomes
\begin{align}
  S_{\rm DP} = \int dt d^3 x \sqrt{-g} \left(
    - \frac{1}{4} g^{\mu\nu} g^{\rho\sigma}
    F_{\mu\rho} F_{\nu\sigma}
    + \frac{1}{2} \Omega^{-2} m^2 g^{\mu\nu} A_\mu A_\nu
  \right).
\end{align}
In the following, we study the production of dark-photon DM based on this action. Considering an isotropic, homogeneous, and spatially flat Universe, the metric can be written as
\begin{align}
  g_{\mu\nu} = \mbox{diag} (1, -a^2, -a^2, -a^2),
\end{align}
where $a$ denotes the scale factor (normalized as $a(t_{\rm now})=1$). 

We assume that the dark-photon mass term is present during and after inflation, so that the effective Lagrangian in Eq.~\eqref{Lagrangian_J} is applicable during the epoch of interest. The simplest way to realize such a mass term is to invoke the Stueckelberg mechanism (for reviews, see, e.g., \cite{Ruegg:2003ps}), which allows one to introduce a bare mass term while preserving gauge invariance for the dark-photon field. In the absence of a specified ultraviolet completion, the kinetic mixing between the dark photon and the electromagnetic photon can then be treated as a free parameter. Alternatively, one may consider a dark Higgs field whose vacuum expectation value generates the dark-photon mass. In this case, the dark Higgs mass (expected to be of the same order as its expectation value) should be larger than the inflationary Hubble scale so that the dark Higgs expectation value is fixed during inflation. If the dark-photon mass is small enough for it to behave as wave-like DM, the associated dark gauge coupling must be extremely small. The size of the kinetic mixing then depends on its origin. In particular, if the kinetic mixing arises from loop effects of particles charged under both the dark and electromagnetic $U(1)$ symmetries, the kinetic-mixing parameter is suppressed by the dark $U(1)$ gauge coupling. As a result, it becomes so small that it would be beyond the reach of future detection experiments. This conclusion can be avoided if one allows for a bare kinetic-mixing term. Since a detailed discussion of the origin of the dark-photon mass is beyond the scope of this work, we simply assume a bare mass term, as can be realized via the Stueckelberg mechanism.

Defining
\begin{align}
  \widehat{m} \equiv \Omega^{-1} m,
\end{align}
the action can be rewritten as
\begin{align}
  S_{\rm DP} = &\,
  \int dt d^3 x \, a^3
  \Bigg[ \frac{1}{2a^2}
    \left\{ (\partial_t \vec{A})^2 + (\vec{\nabla} A_t)^2
    - 2 (\partial_t \vec{A}) \cdot \vec{\nabla} A_t \right\}
    -\frac{1}{2a^4} (\vec{\nabla}\times \vec{A})^2
    \nonumber \\ &
    + \frac{1}{2} \widehat{m}^2
    \left\{
    A_t^2 - \frac{1}{a^2} \vec{A}^2
    \right\}
    \Bigg],
\end{align}
where $\vec{A}\equiv (A_1, A_2, A_3)$. Importantly, there is no term containing the time derivative of $A_t$, so $A_t$ serves as an auxiliary field and can be eliminated via its equation of motion.

To proceed, it is convenient to work in Fourier space. We define the Fourier decomposition by
\begin{align}
  A_\mu (\vec{x}, t) = \int \frac{d^3k}{(2\pi)^3}
  \tilde{A}_\mu (\vec{k},t) e^{-i\vec{k} \cdot \vec{x}},
\end{align}
and, since $A_\mu$ is real, the condition $\tilde{A}_\mu^* (\vec{k}) = \tilde{A}_\mu (-\vec{k})$ holds. In terms of the mode amplitudes, the action reads
\begin{align}
  S_{\rm DP} = & \,
  \int dt \frac{d^3k}{(2\pi)^3} \frac{a^3}{2}
  \Bigg[
    \left( \frac{k^2}{a^2} + \widehat{m}^2 \right)
    \left| \tilde{A}_t - \frac{i \vec{k} \cdot \partial_t \vec{\tilde{A}}}{k^2 + a^2 \widehat{m}^2} \right|^2
    \nonumber \\ &
    + \frac{1}{a^2} \left| \partial_t \vec{\tilde{A}} \right|^2
    - \frac{1}{a^2} \frac{|\vec{k} \cdot \partial_t \vec{\tilde{A}}|^2}{k^2+a^2 \widehat{m}^2}
    - \frac{1}{a^4} \left| \vec{k} \times \vec{\tilde{A}} \right|^2
    - \frac{1}{a^2} \widehat{m}^2 \left| \vec{\tilde{A}} \right|^2
    \Bigg],
\end{align}
where $k\equiv |\vec{k}|$. Varying this action with respect to $\tilde{A}_t$ gives an algebraic equation of motion, which is solved as
\begin{align}
  \tilde{A}_t (\vec{k}) = \frac{i \vec{k} \cdot \partial_t \vec{\tilde{A}} (\vec{k})}{k^2 + a^2 \widehat{m}^2}.
\end{align}
Using this solution, $A_t$ can be eliminated.

For subsequent analysis, we decompose the spatial components $\vec{A}$ into transverse and longitudinal parts:
\begin{align}
  \vec{\tilde{A}} (\vec{k}) =
  \tilde{A}_{T,1} (\vec{k}) \vec{e}_{T,1} (\vec{k})
  + \tilde{A}_{T,2} (\vec{k}) \vec{e}_{T,2} (\vec{k})
  + \tilde{A}_{L} (\vec{k}) \frac{\vec{k}}{k},
\end{align}
where $\vec{e}_{T,\lambda}$ ($\lambda=1,2$) are the polarization vectors for the two transverse modes, satisfying
\begin{align}
  \vec{k} \cdot \vec{e}_{T,\lambda} = 0, ~~~
  |\vec{e}_{T,\lambda}| = 1.
\end{align}
With these, the action splits into the sum of independent transverse and longitudinal contributions,
\begin{align}
  S_{\rm DP} = S_T + S_L,
\end{align}
with
\begin{align}
  S_T = &\,
  \sum_{\lambda=1}^2
  \int dt \frac{d^3k}{(2\pi)^3}a^3 \times
  \frac{1}{2a^2}
  \left[ |\partial_t \tilde{A}_{T,\lambda}|^2 -
    \left( \frac{k^2}{a^2} + \widehat{m}^2 \right) |\tilde{A}_{T,\lambda}|^2
    \right],
  \\
  S_L = &\,
  \int dt \frac{d^3k}{(2\pi)^3}a^3 \times
  \frac{1}{2a^2}
  \left( \frac{a^2 \widehat{m}^2}{k^2 + a^2 \widehat{m}^2} |\partial_t \tilde{A}_L|^2
    - \widehat{m}^2 |\tilde{A}_L|^2
    \right).
  \label{S_L}
\end{align}
We can see that the longitudinal mode $\tilde{A}_L  $ has an inflaton-dependent kinetic function.

It is useful to rewrite the transverse action $S_T$ in terms of the conformal time $\tau$ defined by $d\tau = a^{-1} dt$:
\begin{align}
  S_T = &\,
  \frac{1}{2}
  \sum_{\lambda=1}^2
  \int d\tau \frac{d^3k}{(2\pi)^3}
  \left[ |\partial_\tau \tilde{A}_{T,\lambda}|^2 -
    \left( k^2 + a^2 \widehat{m}^2 \right) |\tilde{A}_{T,\lambda}|^2
    \right].
\end{align}
This expression shows that the transverse modes are approximately conformal, particularly in the relativistic limit $\widehat{m} \ll k/a$. In the strictly massless limit $\widehat{m} \to 0$, $S_T$ becomes independent of the Hubble parameter $H$. As a result, production of transverse modes during inflation is negligible, and they do not contribute significantly to the present DM abundance. Therefore, we will henceforth focus on the longitudinal mode.

Focusing on the longitudinal mode, we first evaluate the quantum fluctuations imprinted on the Fourier amplitude during inflation. The effective mass $\widehat{m}$ depends on the inflaton and generally varies in time. During inflation, however, the inflaton evolves sufficiently slowly that $\widehat{m}$ can be treated as approximately constant on time scales short compared with $H^{-1}$. Introducing
\begin{align}
  \tilde{A}'_L \equiv \frac{\widehat{m}}{k} \tilde{A}_L,
\end{align}
the longitudinal action can be written as
\begin{align}
  S_L \simeq
  \int dt \frac{d^3k}{(2\pi)^3}a^3 \times
  \frac{1}{2}
  \left( \frac{1}{1 + (a^2 \widehat{m}^2/k^2)} |\partial_t \tilde{A}'_L|^2
  - \frac{k^2}{a^2} |\tilde{A}'_L|^2
  \right).
\end{align}
In obtaining this expression, we have neglected terms involving $\partial_t \widehat{m}$, since the above action for $\tilde{A}'_L$ is used only during inflation to evaluate quantum fluctuations. (More rigorous argument will be provided later.) From this form, one sees that at early times in inflation (for $k\gg a\widehat{m}$), $\tilde{A}'_L$ behaves as a massless scalar field. Consequently, $\tilde{A}'_L$ acquires nearly scale-invariant quantum fluctuations during inflation, which in turn determine the present-day DM abundance.

Based on the argument above, we assume
\begin{align}
  \ev{ A_L (\vec{x},t) A_L (\vec{y},t) }
  \xrightarrow{\mathrm{out\, of\, horizon}}
  \mel{0}{\hat{A}_L (\vec{x},t) \hat{A}_L (\vec{y},t)}{0}
  = \int d\ln k \frac{d\Omega_{\vec{k}}}{4\pi}
  \mathcal{P}_{A_L} (k) e^{-i\vec{k} \cdot (\vec{x}-\vec{y})},
\end{align}
where $\ev{\cdots}$ denotes the ensemble (stochastic) average, $\ket{0}$ is the Bunch-Davies vacuum~\cite{Bunch:1978yq}, $\hat{A}_L$ is the corresponding field operator, $\mathcal{P}_{A_L}$ is the power spectrum, and $d\Omega_{\vec{k}}$ is the solid angle element in the direction of $\vec{k}$. 

Then, the power spectrum takes the following form:
\begin{align}
  \mathcal{P}_{A_L} (k) \xrightarrow{k\ll aH}
  \Omega^2 (t_k) \frac{k^2}{m^2}
  \left( \frac{H (t_k)}{2\pi} \right)^2
  \left( 1 + \delta \right),
  \label{P_AL}
\end{align}
where $H$ is the expansion rate and $t_k$ denotes the time at which the mode $k$ exits the horizon. In addition, $\delta$ parameterizes corrections higher order in slow-roll parameters, which arises once we take into account the effect of $\partial_t \widehat{m}$. Notably, in contrast to the original scenario without the Weyl transformation, $\mathcal{P}_{A_L}$ is enhanced by the factor $\Omega^2$, which has a significant impact on the inferred dark-photon mass relevant for the observed DM abundance.

After horizon exit, we assume that the Fourier amplitude is classicalized and that further evolution follows the classical equation of motion. The classical equation of motion is given by
\begin{align}
  \partial_t \left( \frac{\widehat{m}^2}{k^2 + a^2 \widehat{m}^2} \dot{\tilde{A}}_L \right)
  + 3 H \frac{\widehat{m}^2}{k^2 + a^2 \widehat{m}^2} \dot{\tilde{A}}_L 
  + \frac{\widehat{m}^2}{a^2} \tilde{A}_L = 0,
  \label{EoM}
\end{align}
where the ``dot'' is derivative with respect to time $t$. Particularly, for relativistic mode such that $k\gg a\widehat{m}$,
\begin{align}
  \partial_t \left( \widehat{m}^2 \partial_t \tilde{A}_L \right)
  + 3 H \widehat{m}^2 \partial_t \tilde{A}_L
  + \frac{k^2}{a^2} \widehat{m}^2 \tilde{A}_L \simeq 0.
\end{align}
For the out-of-horizon mode, the last term in the above equation is expected to be negligible, and hence $\partial_t \tilde{A}_L\simeq 0$ in such a case, implying that the change of $\tilde{A}_L$ is negligible in the long-wavelength limit. After the reheating, the dark-photon mass settles to $m$, and the superhorizon power spectrum remains as given in~\eqref{P_AL}.

Let us now consider the energy-momentum tensor for the longitudinal mode, focusing on its evolution after inflation. Using the formal definition
\begin{align}
  T_{\mu\nu} =
  2 \frac{\partial \mathcal{L}}{\partial g^{\mu\nu}}
  - \mathcal{L} g_{\mu\nu},
\end{align}
we obtain the local energy density $T_{00}$ as
\begin{align}
  T_{00} =
  \frac{1}{2a^2}
  \left[  (\partial_t A_L) \frac{a^2 m^2}{-\nabla^2 + a^2 m^2} (\partial_t A_L)
    + m^2 A_L^2
    \right].
\end{align}
We are interested in the volume-averaged energy density,
\begin{align}
  \rho \equiv \frac{1}{\mathcal{V}} \int d^3 x T_{00}
  = \frac{1}{\mathcal{V}} \int \frac{d^3 k}{(2\pi)^3}
  \frac{1}{2a^2}
  \left(  \frac{a^2 m^2}{k^2 + a^2 m^2} |\partial_t \tilde{A}_L|^2
  + m^2 |\tilde{A}_L|^2
  \right),
  \label{rho_average}
\end{align}
where the spatial integral runs over a region much larger than the present horizon, and
\begin{align}
  \mathcal{V} \equiv \int d^3 x.
\end{align}

Assuming that the Fourier amplitudes in Eq.\,\eqref{rho_average} can be replaced by their ensemble averages, the energy density can be expressed as
\begin{align}
  \rho (t) \equiv \int d\ln k \, \tilde{\rho} (k,t)
  \label{rho_integral}
\end{align}
with
\begin{align}
  \tilde{\rho} (k,t)
  =
  \frac{d\rho}{d \ln k}
  =
  \frac{1}{2a^2}
  \left( \frac{a^2 m^2}{k^2 + a^2 m^2} \left|\dot{\mathcal{A}}_L (k,t)\right|^2
    + m^2 \left| \mathcal{A}_L (k,t) \right|^2
    \right).
\end{align}
Here, $\mathcal{A}_L$ denotes the mode function satisfying the classical equation of motion; taking $\widehat{m}=m$, Eq.\ \eqref{EoM} gives
\begin{align}
  \partial_t^2 \mathcal{A}_L = 
  - F (t) \partial_t \mathcal{A}_L - \omega^2 (t) \mathcal{A}_L,
  \label{ModeEq.}
\end{align}
where
\begin{align}
  F (t) \equiv &\, \frac{3k^2+a^2(t) m^2}{k^2+a^2(t) m^2} H(t),
  \\
  \omega (t) \equiv &\, \sqrt{ \frac{k^2}{a^2(t)} + m^2 }.
\end{align}
$F$ and $\omega$ describe the Hubble friction and effective oscillation frequency, respectively. The superhorizon initial conditions are given by
\begin{align}
  \dot{\mathcal{A}}_L   \xrightarrow{\mathrm{out\, of\, horizon}} 0,~~~
  \mathcal{A}_L   \xrightarrow{\mathrm{out\, of\, horizon}}
  \mathcal{P}_{A_L}^{1/2} (k).
  \label{superhorizon}
\end{align}
We remark that, after the epoch of radiation-matter equality, structure formation takes place and $\tilde{\rho}$ no longer represents the actual energy spectrum of the dark-photon component. For the purpose of computing the volume-averaged energy density, however, it is sufficient to use $\tilde{\rho}$. During the radiation-dominated epoch when the growth of density fluctuations is negligible, $\tilde{\rho}$ does represent the energy spectrum.

For our later discussion, it is instructive to examine the qualitative behavior of the mode function $\mathcal{A}_L$ and the energy spectrum $\tilde{\rho}$. When $H\gtrsim\omega$, the Hubble-friction term dominates the right-hand side of Eq.~\eqref{ModeEq.}, and $\mathcal{A}_L$ becomes approximately time independent. In this regime, $\tilde{\rho}$ scales as $a^{-2}$. Once the effective oscillation frequency $\omega$ exceeds $H$, the behavior of $\mathcal{A}_L$ changes. For $H \lesssim m\lesssim k/a$, one has $F\simeq 3H$, which implies $\ev{\mathcal{A}_L^2}_{\rm osc}\propto a^{-2}$ (where $\ev{\cdots}_{\rm osc}$ denotes the oscillation average); consequently, $\tilde{\rho}\propto a^{-4}$, as expected for a relativistic fluid. Furthermore, for $H \lesssim k/a\lesssim m$, one finds $F\simeq H$, leading to $\ev{\mathcal{A}_L^2}_{\rm osc}\propto a^{-1}$ and hence $\tilde{\rho}\propto a^{-3}$; in this case, the energy density redshifts as non-relativistic matter.

To determine the spectrum $\tilde{\rho}(k,t)$, which is needed to compute the DM density, we must take into account the evolution of $\tilde{\rho}$ both before and after horizon reentry. Let $k_*$ be the comoving wavenumber of the mode that reenters the horizon when $H=m$. Then, for $H\ll m$, $\tilde{\rho}(k,t)$ scales as $k^2$ for $k\lesssim k_*$, while it scales as $k^{-1}$ for $k\gtrsim k_*$ \cite{Graham:2015rva}. As a result, $\tilde{\rho}(k,t)$ is peaked around $k\sim k_*$. We will confirm this behavior later via numerical calculations.

\section{Starobinsky Inflation and Cosmological History}
\label{sec:starobinsky}
\setcounter{equation}{0}

So far, we have discussed the production mechanism of dark photons in a class of inflationary models in which a Weyl transformation is performed to move to the Einstein frame. Here we consider a specific example of such a model, namely Starobinsky inflation, in order to clarify the importance and relevance of properly incorporating the effects of the Weyl transformation.

\subsection{Starobinsky inflation}

In Strarobinsky inflation model, the action for the gravity sector in the Jordan frame is given by
\begin{align}
  S_{\rm grav} = -\frac{M_{\rm Pl}^2}{2} \int d^4x \sqrt{-\bar{g}}
    \left(\bar{R}-\frac{1}{6m_\phi^2}\bar{R}^2\right),
\end{align}
where $M_{\rm Pl}\simeq 2.4\times 10^{18}\ {\rm GeV}$ is the reduced Planck mass and $m_\phi$ is a mass parameter associated with the inflaton. Here, $\bar{R}$ denotes the Ricci scalar constructed from the Jordan-frame metric $\bar{g}_{\mu\nu}$, and the presence of the $\bar{R}^2$ term indicates that the gravitational sector is non-minimal.

To move to the Einstein frame, it is useful to introduce an auxiliary field $\varphi$, with which the action can be rewritten as
\begin{align}
  S_{\rm grav} = -\frac{M_{\rm Pl}^2}{2} \int d^4x \sqrt{-\bar{g}}
  \left[
    \varphi \bar{R} + \frac{3}{2} m_\phi^2 (\varphi - 1)^2
  \right].
\end{align}
Solving the Euler-Lagrange equation for $\varphi$ yields
\begin{align}
  \varphi=1 - \frac{1}{3 m_\phi^2}\bar{R}.
\end{align}
The Einstein-frame description is obtained by performing a Weyl transformation,
\begin{align}
  g_{\mu\nu} = \varphi \bar{g}_{\mu\nu},
  \label{WeylTr_starobinsky}
\end{align}
which leads to
\begin{align}
  S_{\rm grav} = -\frac{M_{\rm Pl}^2}{2} \int d^4x \sqrt{-g}
  \left[
    R
    - \frac{3}{2} \varphi^{-2} g^{\mu\nu} (\partial_\mu \varphi) (\partial_\nu \varphi)
    + \frac{3}{2} m_\phi^2 (1 - \varphi^{-1})^2
    \right].
\end{align}

In the Einstein frame, the auxiliary field $\varphi$ becomes dynamical. It can be canonically normalized by introducing a scalar field $\phi$ via
\begin{align}
  \varphi = \exp\left( \sqrt{\frac{2}{3}} \frac{\phi}{M_{\rm Pl}} \right).
\end{align}
In terms of the canonically normalized field $\phi$, the action takes the standard form
\begin{align}
  S_{\rm grav} = \int d^4x \sqrt{-g}
  \left[
    -\frac{M_{\rm Pl}^2}{2} R
    + \frac{1}{2} g^{\mu\nu} (\partial_\mu \phi) (\partial_\nu \phi) - V_\phi
    \right],
\end{align}
where the scalar potential is
\begin{align}
  V_\phi = \frac{3m_\phi^2M_{\rm Pl}^2}{4}
  \left[1-\exp\left(-\sqrt{\frac{2}{3}}\frac{\phi}{M_{\rm Pl}}\right)\right]^2.
  \label{V(phi)}
\end{align}
By comparing Eqs.\ \eqref{WeylTr} and \eqref{WeylTr_starobinsky}, we find that the Weyl factor is given by
\begin{align}
  \Omega = \exp\left( \sqrt{\frac{1}{6}} \frac{\phi}{M_{\rm Pl}} \right).
\end{align}
The potential has its minimum at $\phi=0$, around which it can be expanded as
\begin{align}
  V_\phi \simeq \frac{1}{2} m_\phi^2 \phi^2 + \mathcal{O}(\phi^3),
\end{align}
so that $m_\phi$ can be interpreted as the effective mass of $\phi$ around the minimum.

In what follows, we regard the scalar potential in Eq.\ \eqref{V(phi)} as the inflaton potential. For $\phi\rightarrow +\infty$, the potential approaches an asymptotically flat plateau, which enables slow-roll inflation in this regime. The $e$-folding number is related to the inflaton field value as
\begin{align}
  N_e \simeq \frac{3}{4} e^{\sqrt{\frac{2}{3}} \frac{\phi}{M_{\rm Pl}}},
\end{align}
from which one finds
\begin{align}
  \Omega \simeq \sqrt{ \frac{4}{3} N_e }.
  \label{Omega_Ne}
\end{align}
We can also confirm the following approximate expression for the slow-roll parameters:
\begin{align}
  \epsilon \equiv &\,
  \frac{1}{2} M_{\rm Pl}^2 \left( \frac{\partial_\phi V}{V} \right)^2
  \simeq \frac{3}{4} N_e^{-2},
  \label{epsilon_Ne}
  \\
  \eta \equiv &\,
  M_{\rm Pl}^2 \left( \frac{\partial_\phi^2 V}{V} \right)
  \simeq - N_e^{-1}.
\end{align}
Thus, $\epsilon^{1/2}$ and $\eta$ are parametrically of the same order.

During inflation, quantum fluctuations of the dark photon are generated. The computation of the power spectrum is presented in Appendix, and the result is
\begin{align}
  \mathcal{P}_{A_L} (k) \xrightarrow{k\ll aH}
  \Omega^2 (t_k) \frac{k^2}{m^2}
  \left( \frac{H (t_k)}{2\pi} \right)^2
  \left[ 1 +
    \frac{2}{\sqrt{3}} ( 2 - \ln 2 - \gamma_{\rm E} ) \epsilon^{1/2} \right]_{t=t_k},
  \label{P_AL(Starobinsky)}
\end{align}
where $\gamma_{\rm E} \simeq 0.577$ is the Euler-Mascheroni constant. Here, we have included the slow-roll correction up to order $N_e^{-1}$ (i.e., $\mathcal{O}(\epsilon^{1/2})$).

In order to reheat the Universe, the inflaton must decay into Standard Model (SM) fields. One possible inflaton-SM interaction arises from a non-minimal coupling between the Higgs field $H$ and the Ricci scalar:
\begin{align}
  \mathcal{L}_{\rm int} = \xi |H|^2 \bar{R},
\end{align}
where $\xi$ is a coupling constant. With this interaction, the inflaton decay rate into a Higgs-boson pair is \cite{Gorbunov:2012ns}
\begin{align}
  \Gamma_{\phi\rightarrow HH} = \frac{(1-6\xi)^2}{48\pi} \frac{m_\phi^3}{M_{\rm Pl}^2}.
\end{align}
The inflaton can also couple to SM fields through the scale dependence of coupling constants. Since these interactions are loop-suppressed, the corresponding partial decay rates are subdominant when $1-6\xi$ is sizable. 

So far, we have focused on the perturbative decay of the inflaton. We note that if $\xi$ is sizable, the system can undergo so-called preheating~\cite{Kofman:1997yn}, whereby the inflaton energy density is rapidly transferred to the Higgs field soon after the inflaton begins oscillating around the minimum of its potential. The energy density in the Higgs sector is expected to thermalize quickly (assuming that SM Higgs instability at high scales is somehow avoided, e.g., by the presence of additional fields). In this situation, one may effectively estimate $\Gamma_{\phi\to HH}\sim t_{\rm preheating}^{-1}$, where $t_{\rm preheating}\sim \mathcal{O}(10)\,m_\phi^{-1}$ is the characteristic preheating time scale. According to the numerical analysis of Ref.~\cite{Li:2022ugn}, preheating becomes significant for $\xi \gtrsim 1.6$ -- $1.8$. 
For $m_\phi \sim \mathcal{O} (10^{13})\, {\rm GeV}$, the effective decay rate increases abruptly from $\mathcal{O}(10^3)\, {\rm GeV}$ to $\mathcal{O}(0.1)\, m_\phi$ once $\xi$ exceeds this threshold. Thus, the parameter region $\mathcal{O}(10^3)\, {\rm GeV}\ll \Gamma_{\phi\rightarrow {\rm SM}}\ll m_\phi$, where $\Gamma_{\phi\rightarrow {\rm SM}}$ is the decay rate into the SM sector, may not be of primary interest for our purposes. Since $\Gamma_{\phi\to {\rm SM}}$ is a priori unknown and model dependent, we treat it as a free parameter. 

The inflaton can also decay into dark-photon pairs. When the dark-photon mass is much smaller than the inflaton mass, the decay into the longitudinal mode dominates, and the partial decay rate is given by \cite{Li:2021fao}
\begin{align}
  \Gamma_{\phi\rightarrow AA} = \frac{1}{192\pi} \frac{m_\phi^3}{M_{\rm Pl}^2}.
\end{align}
Because the dark-photon field of our interest is very weakly interacting, the decay process into the dark-photon pair does not contribute to the reheating but it causes the production of dark radiation. In our analysis below, we also discuss the constraint from such a production process.

\subsection{Cosmological history}

Now we discuss the evolution of the Universe. We assume that the initial amplitude of $\phi$ is much larger than $M_{\rm Pl}$ so that inflation takes place. The energy density stored in the homogeneous $\phi$ field is
\begin{align}
  \rho_\phi = &\, \frac{1}{2} \dot{\phi}^2 + V_\phi,
\end{align}
and the evolution of the inflaton field is governed by
\begin{align}
  \ddot{\phi} + 3 H \dot{\phi} + \frac{\partial V_\phi}{\partial \phi} =
  - (\Gamma_{\phi\rightarrow {\rm SM}} + \Gamma_{\phi\rightarrow AA}) \dot{\phi}.
  \label{phiddot}
\end{align}

As inflation proceeds, the inflaton amplitude decreases. When $\phi \sim M_{\rm Pl}$, the slow-roll conditions are violated, and the inflaton begins to oscillate about the minimum of its potential. During this oscillatory phase, the inflaton gradually decays into SM particles (as well as into dark photons), thereby reheating the Universe. Reheating is completed when  $H$ drops below the inflaton decay rate.

The evolution of the background radiation in the SM sector is tracked using the entropy density $s$. Assuming rapid thermalization, $s$ obeys
\begin{align}
  \dot{s} + 3 H s = \frac{1}{T} \Gamma_{\phi\rightarrow {\rm SM}} \dot{\phi}^2,
  \label{sdot}
\end{align}
and is related to the cosmic temperature $T$ through
\begin{align}
  T = \left( \frac{45}{2 \pi^2 g_{*s} (T)} s \right)^{1/3},
\end{align}
where $g_{*s}$ denotes the effective number of relativistic degrees of freedom contributing to the entropy density. The radiation energy density is then given by
\begin{align}
  \rho_{\rm r} = \frac{\pi^2}{30} g_* (T) T^4,
\end{align}
where $g_*$ is the effective number of relativistic degrees of freedom contributing to the energy density. In our calculation, we use $g_*$ and $g_{*s}$ provided in Ref.\ \cite{Saikawa:2018rcs}. We also include the contribution from dark photons produced by inflaton decay. We focus on the parameter region in which dark photons from inflaton decay remain relativistic until today. In this case, the energy density of these dark photons (which we refer to as dark radiation) evolves as
\begin{align}
  \dot{\rho}_{\rm DR} + 4 H \rho_{\rm DR} = \Gamma_{\phi\rightarrow AA} \dot{\phi}^2.
  \label{dotrhoDR}
\end{align}
Then, the expansion rate is determined by
\begin{align}
  H = \sqrt{\frac{\rho_\phi + \rho_{\rm r}+ \rho_{\rm DR}}{3 M_{\rm Pl}^2}}.
\end{align}

We note that, when $H\ll m_\phi$, the inflaton oscillates so rapidly that a direct numerical integration of Eq.\ \eqref{phiddot} becomes challenging. In this regime, using the fact that the inflaton motion is governed by a potential dominated by its quadratic term, the inflaton can be well approximated as a pressureless fluid. In this case, the inflaton can be described as a fluid whose energy density (approximately) obeys
\begin{align}
  \dot{\rho}_\phi + 3 H \rho_\phi =
  - (\Gamma_{\phi\rightarrow {\rm SM}} + \Gamma_{\phi\rightarrow AA}) \rho_\phi.
  \label{dotrhophi}
\end{align}
In our numerical analysis, we adopt the fluid description of the inflaton when $H\ll m_\phi$, consistently matching the two descriptions. In this regime, $\dot{\phi}^2$ in Eqs.\ \eqref{sdot} and \eqref{dotrhoDR} is replaced by $\rho_\phi$.

To study the cosmic history, we numerically solve the evolution equations for $\phi$ (or $\rho_\phi$), $s$, and $\rho_{\rm DR}$ from a deep inflationary epoch to a time well after reheating. To identify the parameter region consistent with cosmological observations, we also evaluate the scalar amplitude $A_{\rm s}$ and the scalar spectral index $n_{\rm s}$ as \cite{Stewart:1993bc}
\begin{align}
  A_{\rm s} = &\,
  \left[
  \frac{H^4}{4\pi^2 \dot{\phi}^2}
  \left\{
  1 + (6\epsilon - 2\eta) (2-\ln 2 - \gamma_{\rm E}) - 2 \epsilon
  \right\}
  \right]_{t=t_{\rm pivot}},
  \\
  n_{\rm s} = &\, [1 - 6\epsilon + 2 \eta]_{t=t_{\rm pivot}},
\end{align}
where $t_{\rm pivot}$ denotes the cosmic time at which the pivot scale ($k_{\rm pivot}=0.05\ {\rm Mpc}^{-1}$) exits the horizon.

We then use the resulting background evolution to compute the present-day energy density of dark-photon DM. The evolution of the mode function is governed by Eq.\ \eqref{ModeEq.}. When solving Eq.\ \eqref{ModeEq.}, it is important to note that $F(t)\sim \mathcal{O}(H)$. Consequently, the mode function $\mathcal{A}$ is slowly rolling when $H\gtrsim\omega$. By contrast, when $H\lesssim\omega$, $\mathcal{A}$ oscillates with its amplitude adiabatically decreasing due to cosmic expansion. In principle, even for $H\lesssim\omega$, one may directly use Eq.\ \eqref{ModeEq.} to follow the evolution of the mode amplitude. In practice, however, such a straightforward computation is numerically expensive. To reduce the computational cost, we use the fact that, when $H\ll\omega$, $\mathcal{A}$ oscillates with a frequency approximately given by $\omega$. In this regime, it is convenient to define
\begin{align}
  \mathcal{I}(t) \equiv \frac{1}{2} \dot{\mathcal{A}}^2 + \frac{1}{2} \omega^2 \mathcal{A}^2.
\end{align}
The evolution equation for $\mathcal{I}$ is then found to be
\begin{align}
  \dot{\mathcal{I}} = \left( -F + \frac{\dot{\omega}}{\omega} \right) \mathcal{I},
\end{align}
where we have neglected rapidly oscillating terms. This yields
\begin{align}
  \mathcal{I} (t_{\rm now}) =
  \mathcal{I} (t_0)
  \left( \frac{a (t_{\rm now})}{a (t_0)} \right)^{-4}
  \left( \frac{k^2 + m^2 a^2(t_{\rm now})}{k^2 + m^2 a^2 (t_0)} \right)^{3/2},
\end{align}
where $t_0$ denotes a reference time at which $H\ll\omega$. In our numerical calculations, we use the $\mathcal{I}$-based description for the epoch with $H\ll\omega$, and switch between the two approximations at a sufficiently late epoch so the numerical error in the resulting DM density parameter remains below $1\%$. Notice that, because we are interested in modes which are non-relativistic in the present Universe, $\mathcal{I}$ is related to the energy density spectrum as
\begin{align}
  \tilde{\rho} (t_{\rm now}) = \mathcal{I} (t_{\rm now}),
\end{align}
and hence the DM abundance in the present Universe can be directly calculated from $\mathcal{I} (t_{\rm now})$. 

We also calculate the energy density of the dark-photon component which behaves as dark radiation. Using $\rho_{\rm DR}$, we define
\begin{align}
\Delta N_{\rm eff} \equiv
\left[
\frac{7\pi^2}{120} T_\nu^4 (t_{\rm now})
\right]^{-1}
\rho_{\rm DR} (t_{\rm now}),
\end{align}
where $T_\nu=(4/11)^{1/3} T_\gamma (t_{\rm now})$ is the present neutrino temperature (with $T_\gamma (t_{\rm now})$ being the present temperature of cosmic microwave background).

\section{Numerical Results}

We are now in a position to numerically evaluate the abundance of dark-photon DM generated from quantum fluctuations during Starobinsky inflation. We first determine the background evolution by solving the coupled evolution equations for the inflaton, SM radiation, and dark radiation, given in Eqs.\ \eqref{phiddot} (or \eqref{dotrhophi}), \eqref{sdot}, and \eqref{dotrhoDR}. The initial value of $\phi$ is taken to be much larger than $M_{\rm Pl}$. After obtaining the background evolution, we solve the mode equation given in Eq.\ \eqref{ModeEq.} with the initial conditions specified in Eq.\ \eqref{superhorizon} and calculate the present-day dark-photon abundance using Eq.\ \eqref{rho_integral}.

\begin{figure}[t]
  \centering
  \includegraphics[width=0.7\linewidth]{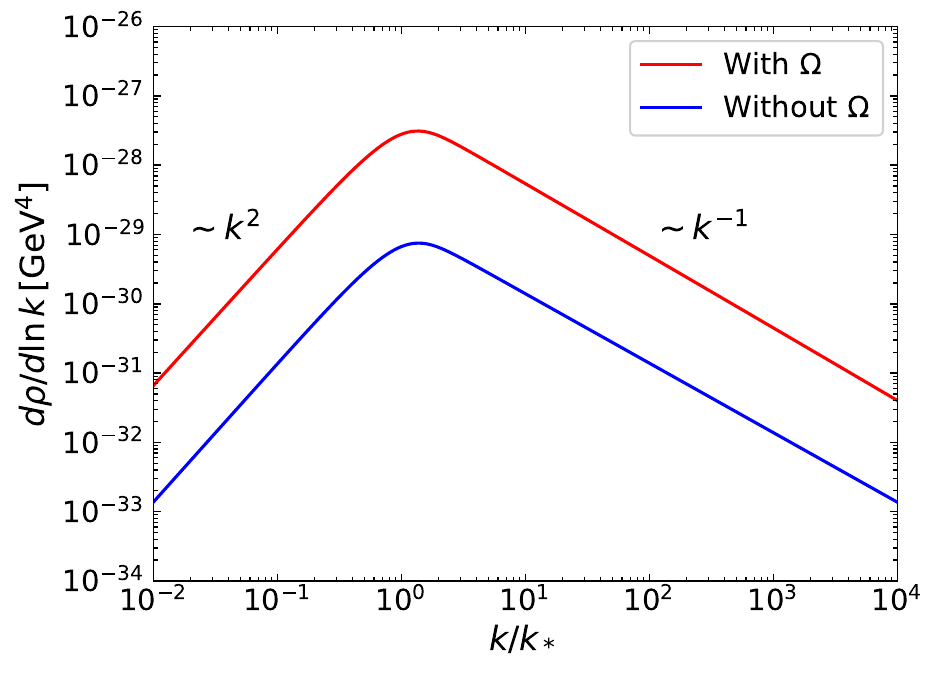}
  \caption{Energy density spectrum of the dark photon, $d\rho/d\ln k$, at $T = 1\, \mathrm{keV}$ for $m_\phi = 3 \times 10^{13}\, \mathrm{GeV}, \Gamma_{\phi\rightarrow{\rm SM}} = 10^3\, \mathrm{GeV}$, and $m = 7\, \mu\mathrm{eV}$. The red curve includes the effect of the Weyl factor $\Omega$, whereas the blue curve is obtained by neglecting this effect. $k_*$ denotes the wavenumber of the mode that reenters the horizon at $H = m$, and $k_* \simeq 1.2 \times 10^3\,\mathrm{pc}^{-1}$ for this case.}
  \label{fig:spectrum}
\end{figure}

Fig.\ \ref{fig:spectrum} shows the energy density spectrum of the dark photon, $\tilde{\rho}=d\rho/d\ln k$, at $T = 1\, \mathrm{keV}$. Here, we take $m_\phi = 3 \times 10^{13}\, \mathrm{GeV}, \Gamma_{\phi\rightarrow{\rm SM}} = 10^3\, \mathrm{GeV}$, and $m = 7\, \mu\mathrm{eV}$. The red curve represents the calculation including the effect of the Weyl factor $\Omega$, whereas the blue curve, shown for comparison, represents the calculation with this effect neglected. As we have discussed, the spectrum including the Weyl factor has a peak at $k \simeq k_*$ and scales as $k^2$ for $k \lesssim k_*$ and as $k^{-1}$ for $k \gtrsim k_*$, as in the original scenario of Ref.\ \cite{Graham:2015rva}. However, the energy density at each wavenumber is enhanced by a factor of $\Omega^2 (t_k) \simeq 40$; note that $\Omega (t_k)$ is nearly constant over a wide range of wavenumbers because of the slow-roll evolution of the inflaton.

The present dark-photon abundance is given by integrating the spectrum. The integration is dominated by $k\sim k_*$. The density parameter is approximately given by
\begin{align}
  \Omega_{\rm DM} h^2 \simeq 0.15 \times 
  \left(\frac{m}{10\, \mu\mathrm{eV}}\right)^{1/2} 
  \left(\frac{\Omega (t_{k_*})}{10}\right)^2 
  \left(\frac{H (t_{k_*})}{10^{13}\,\mathrm{GeV}}\right)^2,
\end{align}
as long as the inflaton dominantly decays into the SM sector.

\begin{figure}[t]
    \centering
    \includegraphics[width=0.7\linewidth]{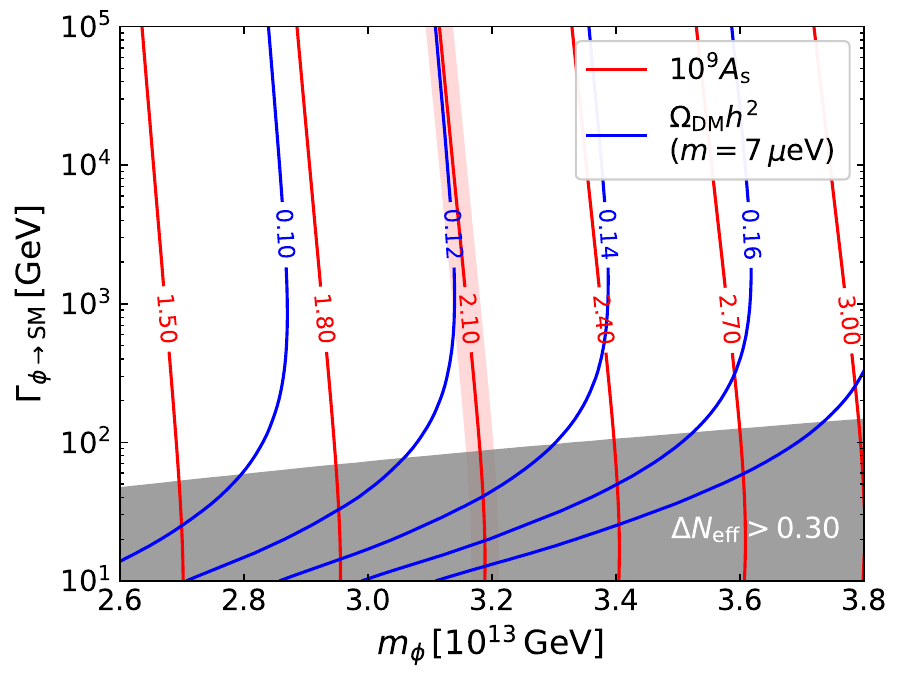}
    \caption{Contours of the scalar amplitude $A_{\rm s}$ and the present dark-photon abundance $\Omega_{\rm DM} h^2$ in the $m_\phi$ vs.\ $\Gamma_{\phi\rightarrow{\rm SM}}$ plane. The dark-photon mass is fixed at $m = 7\, \mu \mathrm{eV}$. The red shaded band shows the region consistent with the observational constraint (i.e., $A_{\rm s}$ is within $(2.100\pm 0.030) \times 10^{-9}$ \cite{Planck:2018vyg}). The gray shaded region ($\Delta N_{\rm eff} > 0.30$) is excluded due to the overproduction of the dark radiation.}
    \label{fig:Omega_contour}
\end{figure}

Fig.\ \ref{fig:Omega_contour} shows the contours of the scalar amplitude $A_{\rm s}$ and the present dark-photon abundance $\Omega_{\rm DM} h^2$ in the $m_\phi$ vs.\ $\Gamma_{\phi\rightarrow{\rm SM}}$ plane. (In the figure, we also consider the region $\Gamma_{\phi\rightarrow{\rm SM}}\gg \mathcal{O}(10^3)\ {\rm GeV}$ to illustrate the asymptotic behavior, although part of this region may be out of our primary interest.) Here, the dark-photon mass is fixed at $m = 7\, \mu \mathrm{eV}$. We have verified that the spectral index $n_{\rm s}$ remains consistent with the Planck constraint, $n_{\rm s} = 0.9649 \pm 0.0042$ \cite{Planck:2018vyg}, throughout the parameter region shown in Fig.\ \ref{fig:Omega_contour}. The parameter region that yields a scalar amplitude consistent with cosmological data (i.e., $A_{\rm s}$ within $(2.100\pm 0.030) \times 10^{-9}$ \cite{Planck:2018vyg}) is indicated by the red shaded band. 

In the same figure, we also show the constraint from the dark-radiation abundance; the gray shaded region conflicts with the constraint $\Delta N_{\rm eff}< 0.30$ (95\% CL) \cite{Planck:2018vyg} and is therefore excluded. Because the partial decay rate into dark photons is sizable, dark radiation can become overabundant, particularly when $\Gamma_{\phi\rightarrow{\rm SM}}$ is relatively suppressed. By contrast, for $\Gamma_{\phi\rightarrow{\rm SM}} \gg \Gamma_{\phi\rightarrow AA}$, the inflaton decays predominantly into the SM sector, and $\Delta N_{\rm eff}$ satisfies the observational bound.

In most of the parameter space consistent with the bound on $\Delta N_{\rm eff}$, the contours of $A_{\rm s}$ and $\Omega_{\rm DM} h^2$ appear nearly parallel. This apparent alignment is, however, accidental. To clarify its origin, it is useful to rewrite the parameter dependence of $A_{\rm s}$ and $\Omega_{\rm DM} h^2$ as
\begin{align}
  A_{\rm s} &\propto \epsilon^{-1} (t_{\rm pivot}) H^2 (t_{\rm pivot}) \sim N_e^2 (t_{\rm pivot}) m_\phi^2, \\
  \Omega_{\rm DM} h^2 &\propto \Omega^2 (t_{k_*}) H^2 (t_{k_*}) m^{1/2} \sim N_e (t_{k_*}) m_\phi^2 m^{1/2},
\end{align}
where we have used Eqs.\ \eqref{Omega_Ne} and \eqref{epsilon_Ne}. When $m_\phi$ and $\Gamma_{\phi\rightarrow{\rm SM}}$ are varied, $N_e (t_{\rm pivot})$ and $N_e (t_{k_*})$ change by nearly the same absolute amount. However, since $N_e (t_{k_*})$ happens to be approximately half of $N_e (t_{\rm pivot})$ ($N_e (t_{\rm pivot}) \simeq 55$, and $N_e (t_{k_*}) \simeq 30$), the relative change in $N_e (t_{k_*})$ is approximately twice that in $N_e (t_{\rm pivot})$. Consequently, $N_e^2 (t_{\rm pivot})$ and $N_e (t_{k_*})$ exhibit nearly the same dependence on $m_\phi$ and $\Gamma_{\phi\rightarrow{\rm SM}}$. It then follows that $\Omega_{\rm DM} h^2 \propto A_{\rm s} m^{1/2}$, so that the contours of $A_{\rm s}$ and $\Omega_{\rm DM} h^2$ are nearly parallel for fixed $m$.

\begin{figure}[t]
  \centering
  \includegraphics[width=0.7\linewidth]{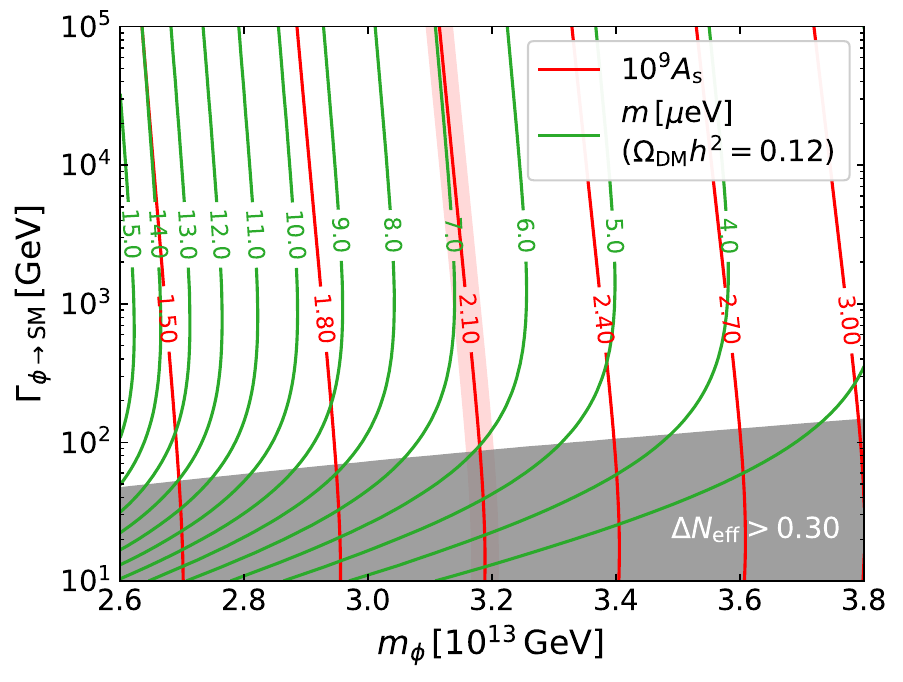}
  \caption{Contours of the scalar amplitude $A_{\rm s}$ and the dark-photon mass $m$ in the $m_\phi$ vs.\ $\Gamma_{\phi\rightarrow{\rm SM}}$ plane. The dark-photon mass $m$ is fixed by imposing $\Omega_{\rm DM} h^2 = 0.12$. The red shaded band shows the region consistent with the observational constraint (i.e., $A_{\rm s}$ is within $(2.100\pm 0.030) \times 10^{-9}$ \cite{Planck:2018vyg}). The gray shaded region ($\Delta N_{\rm eff} > 0.30$) is excluded due to the overproduction of the dark radiation.}
  \label{fig:m_contour}
\end{figure}

To identify the dark-photon mass consistent with cosmological observations, Fig.\ \ref{fig:m_contour} displays contours of the scalar amplitude $A_{\rm s}$ and the dark-photon mass $m$ in the $m_\phi$ vs.\ $\Gamma_{\phi\rightarrow{\rm SM}}$ plane. For each point in this plane, the value of $m$ is fixed by imposing $\Omega_{\rm DM} h^2 = 0.12$. As discussed above, for $\Gamma_{\phi\rightarrow{\rm SM}} \gg \Gamma_{\phi\rightarrow AA}$, we have $\Omega_{\rm DM} h^2 \propto A_{\rm s} m^{1/2}$, and hence $m$ is determined once $A_{\rm s}$ and $\Omega_{\rm DM} h^2$ are specified. 

We find that, imposing the observational constraints $|\Omega_{\rm DM} h^2  - 0.1200| < 0.0012$, $|A_{\rm s} \times 10^9 - 2.100| < 0.030$, and $\Delta N_{\rm eff} < 0.30$ \cite{Planck:2018vyg}, the dark-photon mass should lie in the range $5.6 < m < 7.4\ \mu{\rm eV}$, which corresponds to the oscillation frequency of $1.4 < f < 1.8\ {\rm GHz}$. We note that this conclusion is valid even in the presence of preheating (i.e., for extremely large $\Gamma_{\phi\rightarrow{\rm SM}}$), because the contours of constant $A_{\rm s}$ and $\Omega_{\rm DM} h^2$ are (nearly) parallel.

Interestingly, dark-photon DM in this mass range falls within the target sensitivity of haloscope searches using cavities and other platforms (see, e.g., \cite{Berlin:2024pzi}). We also note that including the effect of the Weyl transformation significantly reduces the dark-photon mass that yields the correct DM abundance.

\section{Conclusions and Discussion}
\setcounter{equation}{0}

In this Letter, we have studied the primordial abundance of dark-photon DM generated from quantum fluctuations during inflation. This production mechanism was first proposed in Ref.\ \cite{Graham:2015rva} in the context of a simple single-field inflationary setup. In the present work, we have instead focused on the case in which the (effective) dark-photon mass varies during and after inflation. This time dependence arises from the unavoidable Weyl transformation required to move from the Jordan frame to the Einstein frame. We have shown that, in such a situation, the predicted DM abundance can be significantly modified compared to the simple inflationary case discussed in Ref.\ \cite{Graham:2015rva}. We have analyzed the abundance in detail and formulated a systematic procedure to evaluate the resulting relic density.

We then applied our formalism to one of the most important examples in this class, namely the Starobinsky inflation model, which necessarily involves a Weyl transformation in order to work in the Einstein frame. We performed a precise calculation of the dark-photon DM abundance, incorporating higher-order corrections in the slow-roll expansion at least up to $\mathcal{O}(N_e^{-1})$. By imposing observational constraints from the observed DM density, the amplitude of the primordial curvature perturbations, and $\Delta N_{\rm eff}$, we found that the dark-photon DM mass is predicted to lie in the range $5.6 < m < 7.4\ \mu{\rm eV}$. This mass range is well within the reach of various haloscope searches for dark-photon DM. We emphasize that the DM mass obtained in our analysis is orders of magnitude smaller than the value inferred when the effect of the Weyl transformation is neglected.

In the present analysis, we have focused in particular on the Starobinsky inflation model, which provides an excellent fit to the cosmological parameters reported by the Planck Collaboration~\cite{Planck:2018vyg}. We note, however, that a potential tension has been discussed between the predictions of Starobinsky inflation and recent results from the ACT observations~\cite{AtacamaCosmologyTelescope:2025blo, AtacamaCosmologyTelescope:2025nti}. At the current stage, it is premature to exclude the Starobinsky inflation model solely on the basis of this tension (see, e.g., \cite{Sabogal:2026qvy}). Moreover, analyses based on the Starobinsky framework remain important for inferring the properties of dark-photon DM. Nevertheless, it is also important to take the ACT results seriously and to investigate the dark-photon DM scenario within inflationary frameworks that are more directly suggested by the ACT data. In particular, some such inflation models require a Weyl transformation in order to move to the Einstein frame. In these cases, the arguments and formalism developed in the present analysis are directly applicable; examples include inflation models proposed in Refs.~\cite{Kallosh:2025rni, Gialamas:2025ofz}. In addition, the abundance of dark photons produced from inflationary fluctuations is sensitive to the post-inflationary cosmic history \cite{Ahmed:2020fhc, Kolb:2020fwh}. For this reason, dedicated studies of the dark-photon abundance in non-standard scenarios are also well motivated. Detailed investigations of these possibilities will be presented elsewhere~\cite{KasamakiMoroi:InPrep}.

\section*{Acknowledgements}

The work of TK was supported by IGPEES, WINGS Program, the University of Tokyo. The work of TM was supported by JSPS KAKENHI Grant No.~23K22486. 

\appendix

\section*{Appendix: Dark Photon Fluctuations}
\setcounter{equation}{0}
\renewcommand{\theequation}{A.\arabic{equation}}

In this appendix, we discuss quantum fluctuations of the longitudinal component of the dark-photon field, following Ref.\ \cite{Stewart:1993bc}. Since we are interested in modes that remain relativistic during inflation, we consider the following action:
\begin{align}
  S_L = &\,
  \int dt \frac{d^3k}{(2\pi)^3}a^3 \times
  \frac{1}{2a^2}
  \left( \frac{a^2 \widehat{m}^2}{k^2} |\partial_t \tilde{A}_L|^2
  - \widehat{m}^2 |\tilde{A}_L|^2
  \right).
\end{align}

Let us introduce
\begin{align}
  \tilde{\pi} (k,t) \equiv
  \frac{m \Omega^{-1}(t_k)}{k} \tilde{A}_L(k,t),
\end{align}
where $t_k$ denotes the epoch at which the mode $k$ exits the horizon. In terms of $\tilde{\pi}$, the action is simplified to
\begin{align}
  S_L = &\,
  \frac{1}{2} \int d\tau \frac{d^3k}{(2\pi)^3}
  a^2 \left( \frac{\Omega}{\Omega(t_k)} \right)^{-2}
  \left[ |\partial_\tau \tilde{\pi}|^2 - k^2 |\tilde{\pi}|^2 \right],
\end{align}
where $\tau$ is the conformal time, defined by $d\tau=a^{-1}dt$.

Next, we define the mode function $u$ by
\begin{align}
  u (k, t) \equiv
  a (t) \left( \frac{\Omega (t)}{\Omega(t_k)} \right)^{-1} \tilde{\pi} (k,t),
\end{align}
with which the action becomes
\begin{align}
  S_L = &\,
  \frac{1}{2} \int d\tau \frac{d^3k}{(2\pi)^3}
  \left[ |\partial_\tau u|^2 -
    \left( k^2 - \frac{\partial_\tau^2 \alpha}{\alpha} \right) |u|^2
    \right],
\end{align}
where
\begin{align}
  \alpha (k,t) \equiv a (t) \left( \frac{\Omega(t)}{\Omega(t_k)} \right)^{-1}.
\end{align}
We can see that $u$ is canonically normalized. Assuming Bunch-Davies vacuum, the mode function $u$ should satisfy (up to an irrelevant phase)
\begin{align}
  u (k,t) \xrightarrow{\tau\rightarrow -\infty}
  \frac{1}{\sqrt{2k}} e^{-ik\tau}.
  \label{initial_u}
\end{align}
In addition, $u\propto \alpha$ at $t\rightarrow\infty$. 

Here, we compute the power spectrum up to $\mathcal{O}(N_e^{-1})$, and thus we neglect terms of order $\epsilon$ (but not $\epsilon^{1/2}$) wherever it yields subleading corrections. Then, we can ignore the time dependence of the expansion rate $H$. The scale factor and the conformal time $\tau$ can be written as
\begin{align}
  a \propto e^{Ht}, ~~~
  \tau \simeq -\frac{1}{aH},
\end{align}
where we adopt the convention $\tau(t\rightarrow\infty)=0$. Furthermore, we obtain
\begin{align}
  \frac{\partial_\tau^2 \alpha}{\alpha} \simeq
  \frac{2}{\tau^2} \left( 1 + \frac{\sqrt{3\epsilon}}{2} \right).
\end{align}
Accordingly, the mode function $u$ obeys
\begin{align}
  \partial_\tau^2 u +
  \left[ k^2 - \left( \nu^2 - \frac{1}{4} \right) \frac{1}{\tau^2} \right] u = 0,
  \label{u-equation}
\end{align}
with
\begin{align}
  \nu \simeq \frac{3}{2} + \sqrt{\frac{\epsilon}{3}}.
\end{align}
The general solution to Eq.\ \eqref{u-equation} is given in terms of the Hankel functions $H_\nu^{(1)}$ and $H_\nu^{(2)}$. Imposing the condition given in Eq.\ \eqref{initial_u}, we obtain
\begin{align}
  u (k,t) = \frac{i}{\sqrt{2k}} \,
  \sqrt{\frac{1}{2}\pi (-k \tau)} \, H_\nu^{(1)} (-k\tau).
\end{align}
We then find the late-time asymptotic behavior
\begin{align}
  u (k,t) \xrightarrow{\tau\rightarrow 0}
  \frac{\alpha (k,t)}{\sqrt{2k}\, a(t_k)} \,
  \left[ 1 +
    \frac{1}{\sqrt{3}} ( 2 - \ln 2 - \gamma_{\rm E} ) \epsilon^{1/2} \right],
\end{align}
which gives
\begin{align}
  \tilde{\pi} (k,t) \xrightarrow{\tau\rightarrow 0}
  \frac{H (t_k)}{\sqrt{2k^3}} \,
  \left[ 1 +
    \frac{1}{\sqrt{3}} ( 2 - \ln 2 - \gamma_{\rm E} ) \epsilon^{1/2} \right].
\end{align}
Using the above expression, we obtain Eq.\ \eqref{P_AL(Starobinsky)}.

\bibliographystyle{RefStyle}
\bibliography{refs}

\end{document}